\documentclass{aa}
\usepackage{graphicx}
\usepackage{txfonts}
\begin{document}
   \title{The distant galaxy cluster CL0016+16: X-ray analysis up to $R_{200}$}


   \author{L. Solovyeva
          \inst{1},
           S. Anokhin \inst{1}, J.L. Sauvageot\inst{1},
           R.Teyssier\inst{1} and D.Neumann\inst{1}
          }

   \offprints{L. Solovyeva,
           \\ \email{lilia.solovyeva@cea.fr}}

   \institute{CEA/DSM/DAPNIA, Service d'Astrophysique, L'Orme des Merisiers, Bat.709,
     91191 Gif-sur-Yvette, France \\
             }

   \date{Received May 15, 2007; accepted , }

\abstract
 {}
{To study the mass distribution of galaxy clusters up to their Virial radius, CL0016+16 seems to be a good candidate,
since it is a bright massive cluster, previously considered as being dynamically relaxed. }
{Using XMM-Newton observations of CL0016+16, we performed a careful X-ray background analysis, and,
we detected convincingly its X-ray emission up to $R_{200}$. We then studied its dynamical state with
a detailed 2D temperature and surface brightness analysis of the inner part of the cluster. }
{Using the assumption of both spherical symmetry and hydrostatic equilibrium (HE) we can determine the main cluster parameters: total
mass, temperature profile, surface brightness profile and $\beta$-parameter. We also build a temperature map which clearly exhibits
departure from spherical symmetry in the centre. To estimate the influence of these perturbations onto our total mass estimate, we also
compute the total mass in the framework of the HE approach, but this time with various temperature profiles obtained in different directions.
These various total mass estimates are consistent with each other. The temperature perturbations are clear signatures of ongoing merger activity.
We also find significant residuals after subtracting the emissivity map by a 2D $\beta$-model fit. We conclude that, although
CL0016+16 shows clear signs of merger activity and departure from spherical symmetry in the centre, its X-ray emissivity can be detected
up to $R_{200}$ and the corresponding mass $M_{200}$ can be computed directly.
It is therefore a good candidate to study cosmological scaling laws as predicted by the theory.}
{}

 \keywords{galaxies:cluster:individual: CL0016+16: observation-X-rays }
\authorrunning{L. Solovyeva et al.}
\titlerunning{The distant galaxy cluster CL0016+16: X-ray analysis up to $R_{200}$}
\maketitle
%

\section{Introduction}

In the hierarchical scenario of structure formation, clusters of galaxies are the largest and youngest {\it virialized} objects in the Universe.
This makes them ideal targets for cosmological studies. Clusters of galaxies are self-similar in shape, and the cluster population obeys scaling laws
for various physical properties~: total mass, temperature and luminosity. The evolution of these properties with redshift gives also complementary
information that shed light on cluster physics.
X-ray observations of galaxy clusters allow us to study the hot intra cluster medium (ICM), which is the main baryon reservoir in galaxy clusters.
In the era of XMM-Newton and Chandra observations, we can obtain detailed information on the density and temperature
distribution of the ICM, and study their internal structure with unprecedented accuracy.

This paper reports a detailed study of CL0016+16, a very massive, luminous and distant (z = 0.54) galaxy cluster. CL0016+16 is one
of the most extensively studied cluster of galaxies in different wavelengths, in particular in the optical \citep{Tanaka}, radio \citep{Feretti}
bands and for Sunyaev-Zel'dovich effect \citep{Birkinshaw}, \citep{Bonamente}. It has been also studied using weak lensing \citet{Clowe} and X-rays from ROSAT \citep{Neumann} and XMM-Newton \citep{Kotov},
\citep{Worrall}.

The question about the dynamical state of this cluster is still open.  Previous X-ray analysis concluded that the cluster might be relaxed
\citep{Kotov}. But we know that this cluster has a radio halo from \citet{Feretti}, which can be a signature of merger activity.
Since the total mass is computed using the hydrostatic equilibrium approach, we need to have a better idea on the dynamical state of the
cluster.

In this paper, we present a detailed spectro-imaging study of the galaxy cluster CL0016+16. We first assumed that the cluster is relaxed and
spherically symmetric. We obtained high quality surface brightness profiles using different backgrounds subtraction methods and demonstrated
that we can detect the cluster emission up to $R_{200}$. This specific radius was first used by theoreticians to define dark matter halos, for
which the mean density is 200 times the critical density of the universe. N body and hydrodynamics numerical simulation have shown that
within this radius, the gas and dark matter halo can be reasonably considered as being in dynamical  equilibrium \citep{Cole}. To compute
$R_{200}$, we use the standard definition:
\begin{equation}
\label{density_cont}
M_{200}=\frac{4\pi}{3}200\rho_{crit}(z)R^{3}_{200}
\end{equation}
where $\rho_{crit}=3H^{2}_{0}/8\pi G$ is the critical density, our
cosmological parameters $\Omega_{m}=0.3$, $\Omega_{\Lambda}=0.7$ and
$H_{0}$ = 70 km s$^{-1}$Mpc$^{-1}$.
Most theoretical predictions (mass function, density profiles, scaling laws) are computed within this radius, while most X-ray observations
are limited to much inner regions, such as $R_{1000}$. The purpose of this paper is to compute the observed properties of a large X-ray
cluster within $R_{200}$.
Using the hydrostatic equilibrium and first guesses of the $\beta$-model parameters
and mean temperature of 8.9 keV taken from
\citet{Kotov}, we estimated using Eq.\ref{density_cont} that $R_{200}$ = 1.92 Mpc or $5\arcmin$ for z=0.54. We, then verify that our measurement of temperature at  $R_{200}$  is very similar to the used value.
In a second step, we performed a detailed spectral analysis of CL0016+16, and deduce its average temperature, its temperature profile and a
high quality temperature map. To estimate the influence of temperature variations on the total mass estimate, we calculated the mass profile
using the hydrostatic equilibrium equation and different temperature profiles in different directions. Finally, using the $\Lambda CDM$
cosmological model, we have tested how CL0016+16 fits into cosmological self-similarity theory.
All errors on the cluster parameters were obtained at the $68\%$ confidence level.


\section{Data analysis}

We studied the physical parameters and ICM dynamics of CL0016+16 up to $R_{200}$. The cluster emission is weak in the external region,
so the astrophysical background can play the main role in the outer region, that is why the treatment of background is very important.
To obtain the best result we used three different backgrounds for subtraction in the data analysis. These were the background of A.
Read without "flare" rejection, the background of J. Nevalainen with "flare" rejection and the modelling of background using the
observation data.

The method of double background subtraction from \citet{Arnaud} was
used. For analysis we used the XMM- Newton data from EPIC cameras
(MOS 1,2 and pn) and the XMM-Newton Science Analysis System (SAS)
for data reduction. In the MOS 1,2 data set we took into account
event patterns 0 to 12 and in pn data - patterns 0 to 4, flag = 0.

The sky coordinates of background observation in the event files were modified using the aspect solution of CL0016+16 observation.

From count rate of the observation data we detected and excluded the periods of "flare". Using the count rate of the observation in the high energy bands (10-12 keV) and exposure time of the observation we normalized the background.

The effective area of the XMM-Newton mirrors is a function of the off-axis angle and energy of the photons. One could manage it through a weight function directly computed for each event \citep{Majerowicz} or through exposure map.  The vignetting is a geometrical effect, both approaches lead to similar results. All along of this work we have used the weight method.

We excluded all detectable point sources from the data observation in our spectral and spatial analysis. The sources were detected in the 0.3-4.5 keV energy bands. Detected point sources were masked with circles of $70\%$ point spread function power radii.

\subsection{Double subtraction}

After cleaning the flare events, the XMM background is dominated by
the Cosmic X-ray background (CXB) and non X-ray background (NXB)
induced by high energy particles. In our analysis we used the
double-subtraction method by \citet{Arnaud} for background
subtraction.

The background subtraction for profile and spectrum consists of two steps. Firstly we subtract the normalized blank field obtained using the same spatial and energy selection  (NXB component) and then we subtract the residual components using the data in the outer part of the cluster emission (CXB component).

\subsection{Background subtraction (blank field of A. Read) }

The background of A. Read is a blank field without the any "flare" rejection \citep{Read}. We excluded the "flare" period using the similar method for the observation and background. We suggested that the residual "flare" background could play a role in each energy band. That is why for the observation and background data we performed the "flare" rejection in the standard selection \citep{Majerowicz} and for each energy band (10-12 keV, 0.3-12 keV, 2-5 keV, 0.3-2 keV). To determine the best limit of "flare" background we used the light curve, we compared the histograms obtained from the images in the external regions. Using different energy bands for the "flare" rejection we obtained the same results and the same exposure times.

\subsection{Background subtraction (blank field of J. Nevalainen) }

To obtain the best result from the image analysis, to detect the
cluster emission up to $R_{200}$ with the XMM-Newton data we also
performed the image analysis with the background of J. Nevalainen
\citep{Nevalainen}. The background data of J. Nevalainen is cleaned
for "flares", has better statistic, less sources, no artefacts in the
centre of FOV. In this analysis for observation we excluded the
"flare" rejection in the standard selection using the Poisson filter
the flares were detected as $>$ 3$\sigma$ deviation from the mean.

\begin{figure*}[!!!htb]
\begin{center}
\resizebox{\hsize}{!} { \hspace{0mm}
\includegraphics{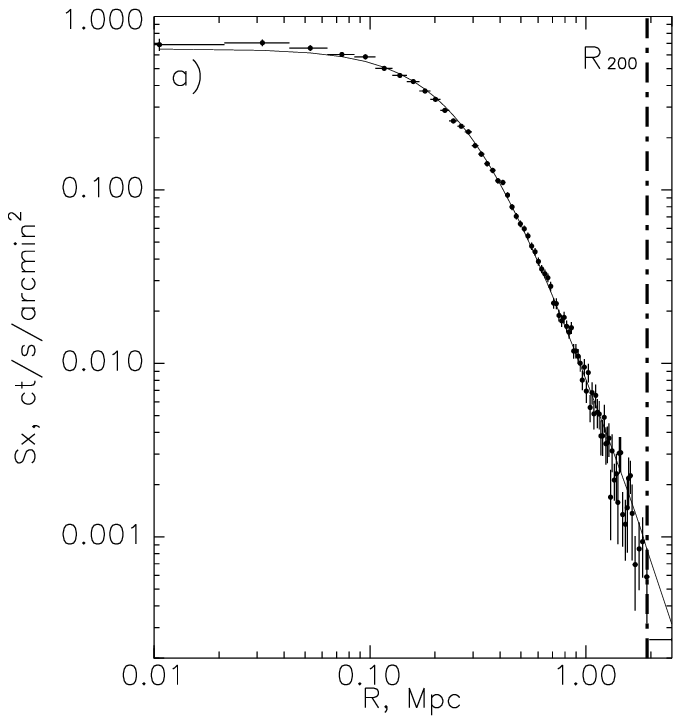}
\hspace{0mm}
\includegraphics{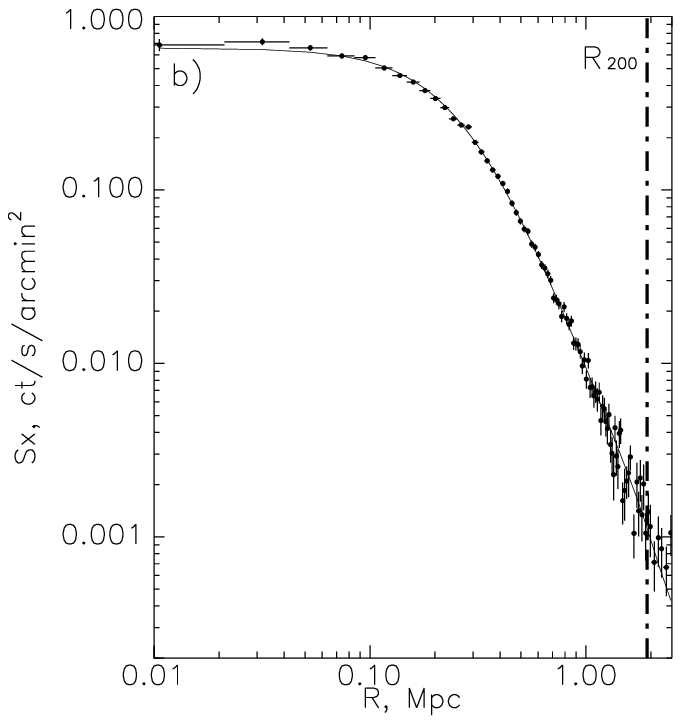}
\hspace{0mm}
\includegraphics{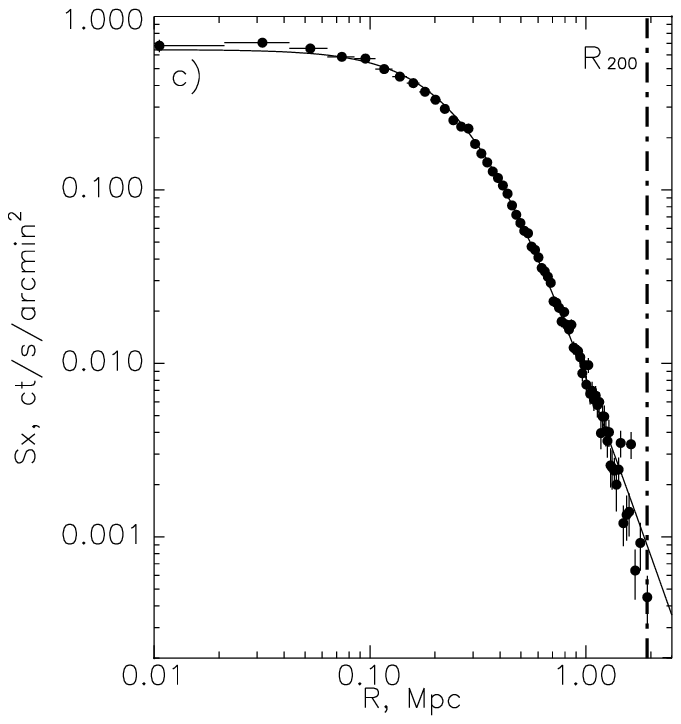}
}
 \caption{\label{SBpr} The surface brightness profiles combined MOS1,MOS2 and pn cameras obtained using different backgrounds a) the background of A. Read, b) the background of J. Nevalainen, c) the background model from observation data. Solid line is the best fit $\beta$-model.}
\end{center}
\end{figure*}

\section{Image analysis}

\subsection{Surface brightness profile}

The surface brightness profile was fitted with the $\beta$-model \citep{Cavaliere} in which the surface brightness $S(r)$ is defined as:

\begin{equation}
S(r)=S_{0}\left(1+\frac{r^{2}}{r_{c}^{2}}\right)^{-3\beta + 0.5}
\end{equation}
where $S_{0}$ is the central intensity, $r_{c}$ is the core radius, $\beta$ is the slope parameter. The best fit $\beta$-model was used to compute the total mass. The $\beta$-model is a model that allows us to project easily the surface brightness of the ICM which emits via thermal bremsstrahlung into the following gas density profile:

\begin{equation}
n(r)=n_{0}\left(1+\frac{r^{2}}{r_{c}^{2}}\right)^{-3\beta/2}
\end{equation}

\subsubsection{Surface brightness profile with blank field of A. Read }

To obtain the surface brightness profile using the background of A. Read we created the images in the 0.3-4.5 keV and calculated the corresponding exposure maps taking into account the detector geometry for the observation and background data for three cameras of XMM-Newton. The radial surface brightness profiles were created from these images and three profiles were summed. Significant cluster emission was detected up to $R_{200}$ with the limit of detection 3$\sigma$. The $\beta$-fits of cluster profile are given in the Table \ref{fit_beta} and the convolved best fit $\beta$-model is plotted in Fig. \ref{SBpr}a.

\subsubsection{Surface brightness profile with blank field of J. Nevalainen }

To obtain surface brightness profile using background of J.
Nevalainen we extracted the surface brightness profile of the
cluster in the 0.3-4.5 keV energy band. This band was chosen to
optimise the signal-to-noise ratio. We binned the photons into
concentric annuli with a size of 1.65$\arcsec$ centred on the
maximum of the X-ray emission for each camera. The three profiles
were then summed. The resulting surface brightness profile $S_{0}$
is shown in Fig. \ref{SBpr}b. The cluster emission is clearly
detected up to $R_{200}$, with limit of the detection 3$\sigma$. We
fitted $S_{0}$ with $\beta$-model, without the first three points
and including the PSF deconvolution (see the results in Table
\ref{fit_beta}). The same results of the $\beta$-model fit parameters
were obtained using the all points for the fits, $\beta$ = (0.73
$\pm$ 0.02) and $r_{c}$ = (0.27 $\pm$ 0.02) Mpc. This result is in a
good agreement with the result of \citet{Kotov}.

\begin{figure}
\begin{center}
\resizebox{\hsize}{!}
    {
     \hspace{0mm}
     \includegraphics{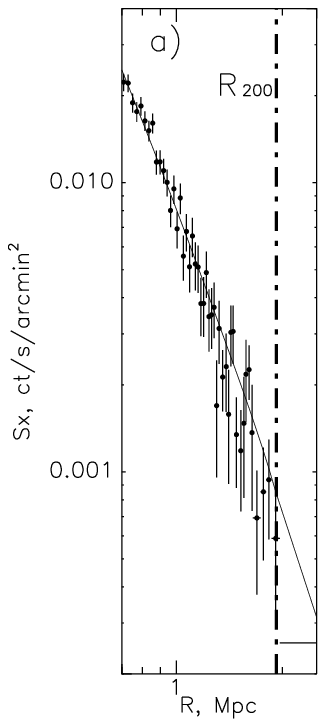}
     \hspace{0mm}
     \includegraphics{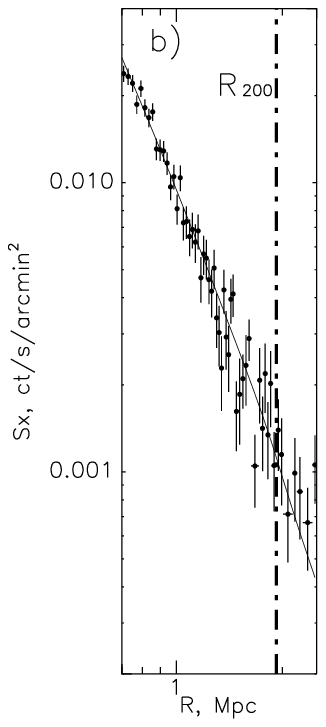}
     \hspace{0mm}
     \includegraphics{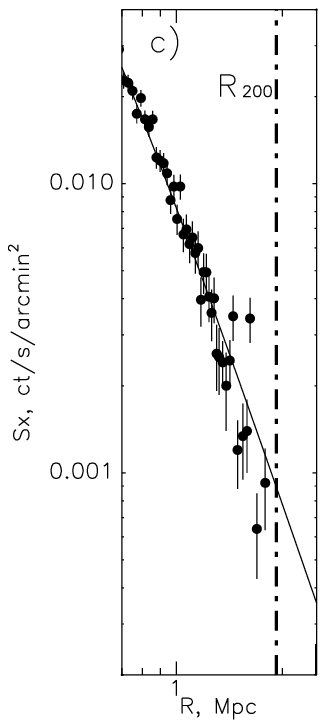}
    }
\caption {\label{SBpr_zoom} The combined (MOS1,MOS2, pn) surface brightness profiles in the external region obtained using different backgrounds a) the background of A. Read, b) the background of J. Nevalainen, c) the background modelling from corresponding observation data. Solid line is the best-fit $\beta$-model.}
\end{center}
\end{figure}

\subsubsection{Surface brightness profile using the modelling of background
from observation data}

The astrophysical background plays a very important role in the data
analysis of the outskirts of galaxy clusters so we assumed that would be
better to use the observation data for the background modelling.
Using observation data we modelled the CXB and NXB components of the
background. The CXB component of background is vignetted by
X-optics, but the NXB component is not \citep{Arnaud}. In this analysis we did the
following: selection of the "flare" events in the observation data,
the surface brightness profile creation from observation data, using
the vignetting function we searched the CXB and NXB components of
the background by minimising $\chi^{2}$ in the outer regions.
Finally we subtracted the CXB and NXB components from the
observation. We obtained cluster emission up to $R_{200}$. After
rebinning the surface brightness profile was fitted with
$\beta$-model, see Fig.\ref{SBpr}c. The fit of the cluster profile
is given in Table \ref{fit_beta}.

\begin{table*}[!!htb]
\caption{The comparison of the $\beta$-model fitting results for the
galaxy cluster CL0016+16. }\label{fit_beta}
\centering
\begin{tabular}{l l l l l l l}
\hline
\hline
 &          $\beta$-fit& (0.01-2.5 Mpc)&& $\beta$-fit by region extern &(0.3-2.5 Mpc) &\\
 \hline
 data source & $\beta$ & $r_{c}$(Mpc) & $\chi^{2}_{reduced}$& $\beta$ & $r_{c}$(Mpc) & $\chi^{2}_{reduced}$\\
\hline
\hline
   Kotov and Vikhlinin 2005 &  0.76 $\pm$ 0.01& 0.27 $\pm$ 0.01& 1.38 & & & \\
   Worrall and Birkinshaw 2003&  0.70 $\pm$ 0.01  & 0.23 $\pm$ 0.01 & 1.35 &  & & \\
   Neumann and Bohringer 1996 & 0.68 $\pm$ 0.3 & 0.28 $\pm$ 0.14& 1.02& & &\\
\hline
    Using  background by A.Read & 0.77 $\pm$ 0.01& 0.31 $\pm$ 0.01& 1.56&  0.80 $\pm$ 0.02& 0.34 $\pm$ 0.02& 1.35\\
    Using  background by J.Nevalainen & 0.76 $\pm$ 0.01& 0.29 $\pm$ 0.01& 1.15& 0.72 $\pm$ 0.01& 0.29 $\pm$ 0.01& 1.02 \\
    Using  the modelling \\of background from observation& 0.76 $\pm$ 0.01& 0.38 $\pm$ 0.01& 1.57&  0.81 $\pm$ 0.02& 0.38 $\pm$ 0.02& 1.21\\
\hline
\end{tabular}
\end{table*}

\subsection{Summary of image analysis }

In our image analysis,
the same results were obtained using different backgrounds for subtraction.
The surface brightness profile of
CL0016+16 was detected up to $R_{200}$, this is
important for studying the cluster physics in the outskirts.

In all samples the $\beta$-model is in a good agreement with the surface brightness profiles using the method of double background subtraction. The different results of the fitting $\beta$-model for CL0016+16 are shown in Table \ref{fit_beta}, a good agreement with other authors was obtained.

We obtained worse reduced $\chi^{2}$ using the background of A. Read and the modelling of background. It may be because background of A. Read has artefacts in the centre of FOV so the surface brightness profile of the cluster does not fit well in the centre with $\beta$-model. Also in this case the variation of reduced $\chi^{2}$ depends on the second background subtraction.

We established that the $\beta$-fit had problems in fitting the core
radius, so we fitted the $\beta$-model with the surface brightness
profile in the external region and obtained the better values of
reduced $\chi^{2}$. Also, using the template of simulated cluster we
tested the possibility of detection up to $R_{200}$ with XMM-Newton
data (see Appendix \ref{appendix} ).

With XMM-Newton data we detect cluster emission of CL0016+16 up to
$R_{200}$ then it was possible to determine the cluster total mass and
physical parameters more precisely, to study physics near the virial
radius and to test self similarity theory. For our spectral analysis
we decided to use the background of Nevalainen, because using this
background we obtained a better result with the $\beta$-model fit.

\section{Spectral analysis}
We needed to study the ICM dynamics of CL0016+16 and to calculate more precisely the total mass profile up to $R_{200}$ using the equilibrium approach.

From the spherical brightness emission, CL0016+16 looks as a relaxed cluster without any cool core. On the other hand CL0016+16 has the companion cluster RX J0018.3+1618 \citep{Worrall} and halo radio \citep{Feretti}. It is very important to check the equilibrium of this system.

To test the equilibrium we performed a detailed spectral analysis.
The mean temperature, temperature profile, temperature map,
temperature in the regions and the temperature profiles in different
directions were obtained. In our spectral analysis we used three
cameras of XMM-Newton, the background from J. Nevalainen and the
method of double background subtraction by \citet{Arnaud}.
We fitted
the spectra with XSPEC using the redshifted APEC plasma emission
model with the absorption $N_{H}$ = $4\cdot 10^{20}$ cm$^{-2}$ and
free abundances. In our research we are more interested in
temperature variations, at same time we fitted temperature with free
and fixed abundance (0.18 solar unit, obtained from the
best fit of the mean temperature in $6.4\arcmin$ radius). For each
spectrum similar temperatures were obtained in both case
but sometimes we cannot determine reliably the
abundances due to weak statistics.

\subsection{Mean temperature}

\begin{figure}[!!htb]
\begin{center}\includegraphics[
   width=5.9cm,
angle=270, trim=45 0 15 0, clip,
   keepaspectratio]{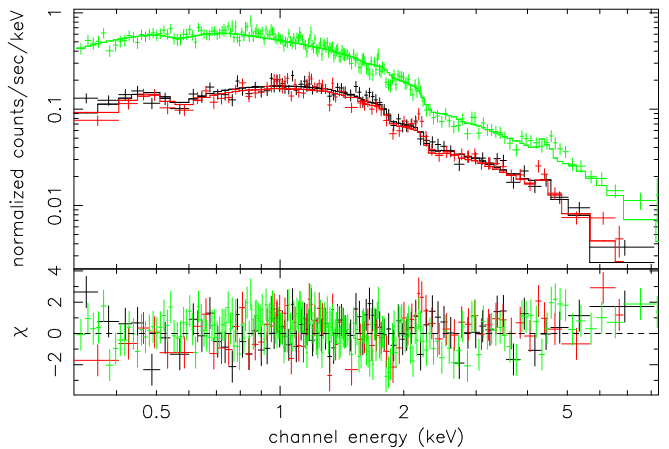}\end{center}
\caption{\label{spectra_amas_all5arcmin} CL0016+16 extracted
spectra, in the region $5\arcmin$, from the XMM-Newton data. Black,
red, green points are the data from the corresponding EPIC/ MOS1,
MOS2 and pn data. The solid lines is the best fit of the isothermal
model with kT = 8.81 keV and abundance of 0.18 solar units. }
\end{figure}

The overall MOS1, MOS2 and pn spectra extracted from the event
file are shown in Fig. \ref{spectra_amas_all5arcmin}. The spectra
were corrected for the vignetting and the background. The
integration region for the cluster was restricted to $5\arcmin$ ($R_{200}$),
this region was chosen to test the self similarity
theory. The temperature values were estimated for all EPIC cameras
MOS1, MOS2 and pn. The best fit gives $kT=8.81\pm0.35$ keV and
an abundance of $0.18\pm0.05$. The reduced $\chi^{2}$ of 0.93.
We also estimated the mean temperature
in $6.4\arcmin$ radius, the best fit gives $kT=8.83\pm0.36$ keV with
reduced $\chi^{2}$ of 0.93 and an abundance of of $0.19\pm0.05$
solar units.

\begin{table*} [!!htb]
\caption{The spectral fits results of the temperature profile with
frozen and free abundances for all EPIC cameras, the abundance fits, the temperature for  MOS1 and MOS2 cameras, pn camera.} \label{Tprof}
\centering
\begin{tabular}{l c c c c c }
\hline\hline annulus & $T (keV)/\chi^{2}_{red}$ & $T
(keV)/\chi^{2}_{red} $ & Z ( $Z_{\odot}$) & $T
(keV)/\chi^{2}_{red}$&
$T (keV)/\chi^{2}_{red} $  \\
& Z = 0.18$Z_{\odot}$ & free Z& & free Z & free Z  \\
$r_{1}$-$r_{2}$($\arcmin$)& mos1+mos2+pn & mos1+mos2+pn &mos1+mos2+pn & mos1+mos2 & pn \\
\hline
&\\
0.0-0.5 &10.63$^{+0.57}_{-0.47}$/0.92 & 10.27$^{+048}_{-0.47}$/0.91 & $0.42\pm0.11$ & 11.45$^{+1.25}_{-0.95}/0.88$ & 9.58$^{+0.62}_{-0.61}/0.93$ \\
0.5-1.0 &9.31$^{+0.42}_{-0.42}$/1.01& 9.34$^{+0.43}_{-0.43}$/1.01&$0.15\pm0.06$& 9.80$^{+0.68}_{-0.67}/1.02$&9.05$^{+0.57}_{-0.56}/1.00$\\
1.0-2.0 &9.17$^{+0.51}_{-0.51}$/0.92 & 9.33$^{+0.55}_{-0.54}$/0.91&$0.11\pm0.07$& 10.23$^{+1.19}_{-0.88}/0.94$&8.77$^{+0.68}_{-0.63}/0.89$ \\
2.0-4.0 &6.45$^{+0.88}_{-0.62}$/1.03& 6.32$^{+0.75}_{-0.61}$/1.03&$0.38\pm0.17$ & 8.06$^{+1.96}_{-1.36}/1.03$&5.19$^{+0.88}_{-0.56}/1.02$ \\
4.0-6.4 &4.14$^{+2.18}_{-1.27}$/0.87& 4.07$^{+2.21}_{-1.30}$/0.85&$ ... $&  1.99$^{+2.43}_{-0.55}/0.89$&4.98$^{+2.49}_{-2.51}/0.82$ \\
&\\
\hline
\end{tabular}
\end{table*}

\subsection{Temperature profile}

For a more exact determination of the total mass profile it is better to use the temperature profile. We supposed that CL0016+16 was a relaxed cluster and assumed that the temperature structure of this cluster was spherically symmetric. The spectra were extracted in the five concentric annuli centred on the cluster X-ray emission peak and we fitted the data as described above. The sizes of the annuli were chosen to optimise the signal-to-noise in each annulus. Note, we fitted spectra for temperature profile with frozen and free abundances  we obtained a similar values of the temperature and error bar (see Table \ref{Tprof}).

\begin{figure}[!!htb]
\begin{center}\includegraphics[
   width=7.6cm,
   keepaspectratio]{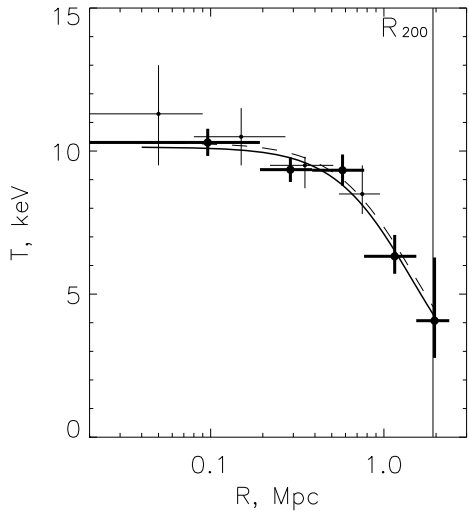}\end{center}
\caption{\label{TprofileOwnKotov}The solid points are the total temperatures best fit, the solid line is
the fitting of the temperature profile with the function from \citet{Kotov},
the thin points are the total temperatures best fit of  \citet{Kotov} and dotted line are the obtained total temperatures profile of \citet{Kotov}.}
\end{figure}

We obtained the total temperature profile up to $R_{200}$ (see Fig.\ref{TprofileOwnKotov}). The temperature profile is in good agreement with spectral fit results of \citet{Kotov}. The solid line profile is that from our analysis and the dotted line profile is the profile by \citet{Kotov}. Note, that temperature was detected up to $R_{200}$ and obtained points are in a good agreement with the function of temperature profile by \citet{Kotov}.

\subsection{Temperature map}

To understand the ICM dynamics of CL0016+16 it is necessary to obtain a temperature map, which is the best accessible measurable indicator of a system equilibrium. The temperature map was obtained by X-ray wavelet spectral mapping algorithm (XWSM).
The spatial temperature variations are coded at different scales in the
wavelet space using the Haar wavelet and denoised by thresholding the wavelet coefficients. For a complete description of the  algorithm,
see  \citet{Bourdin}.

\begin{figure}[!!htb]
\begin{center}\includegraphics[
   width=7.6cm,
   keepaspectratio]{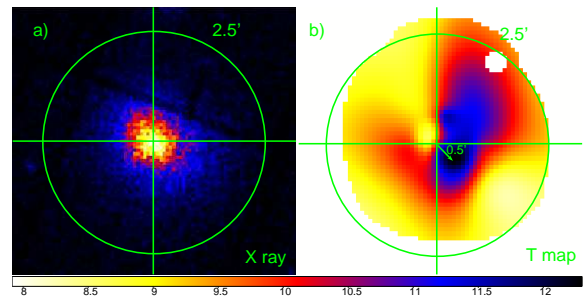}\end{center}
\caption{\label{Xray_TmapXWSM_circle2.5} a) the X-ray emission from
EPIC/MOS1, MOS2 \&pn in the energy band 0.3-4.5 keV. b) the
temperature map obtained by wavelet algorithm described in
\citet{Bourdin}.}
\end{figure}

The CL0016+16 temperature map was computed using three cameras of
XMM-Newton. This algorithm fitted the temperature with the APEC plasma
emission model. The resulting temperature map shown in
Fig.\ref{Xray_TmapXWSM_circle2.5} was obtained from a wavelet
analysis performed on 5 scales corresponding to a structure of
a minimum size of 45$\arcsec$ using the mean temperature of 9 keV and
with the same absorption and abundance.

The temperature map was obtained up to $2.5\arcmin$ (1 Mpc) with a
$68\%$ confidence level for all detected structures. The overall
appearance of the temperature structure is strongly asymmetric. The
maximum of the temperature is 12.52 keV and in the
cold region of the cluster it is 8.7 keV. We observed the maximum
of temperature on the south-west from the cluster emission centre,
and minimum of temperature on north-east from cluster
emission centre. Fig. \ref{Xray_TmapXWSM_circle2.5} shows the hot
regions to south-east and to north-west and two cold regions
in other directions. Note, that the maximum of the temperature
is not superimposed onto the maximum emission. The peak of
the temperature is displaced by the $0.5\arcmin$ (190 kpc) from the
cluster emission peak.

\subsection{Temperature in selected regions}

\begin{figure}[!!htb]
\begin{center}
\includegraphics[width=7.6cm, keepaspectratio]
{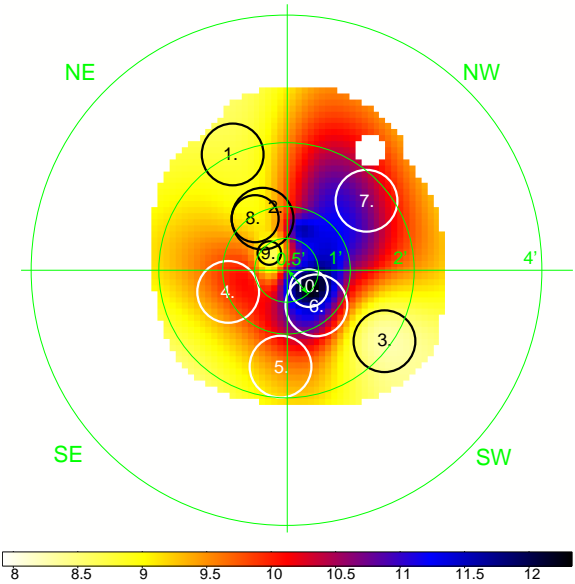}\end{center}
\caption{\label{TmapReg_Panda.ps}The white contours illustrated the chosen regions from which spectra were extracted. The regions were defined on the basis of temperature map. The green contours are the chosen sectors, to extract the spectra for four temperature profiles.}
\end{figure}

To check the temperature map obtained with XWSM we computed the hardness ratio map and we observed the same variation in the cluster centre, in the direction of the temperature map elongation from south-east (SE) to north-west (NW). To assess the significance of the temperature variations found in the temperature map, we focused on the spectral fits in specific regions, see Fig. \ref{TmapReg_Panda.ps}. The regions were chosen from temperature map. These results are shown in Table \ref{Tregions}. The temperature obtained for the regions is in a good agreement with that from spectral wavelet analysis. More interesting results were obtained for the regions 9, 10 and 3, where the temperatures are 7.81$^{+1.01}_{-0.93}$ keV, 12.01$^{+1.45}_{-1.34}$ keV and 6.16$^{+1.14}_{-0.87}$ keV, respectively. The regions 9 and 3 are the temperature minima and the region 10 has the highest of temperature. The temperature peak does not coincide with the maximum of cluster emission. From our spectral analysis we established that the temperature variations are approximately 4 keV in the cluster center. It is important to understand whether these variations of temperature would give any significant effect on the total mass.

\begin{table} [!!htb]
\caption{The spectral fits results in the chosen regions, with free abundances. Three EPIC cameras were used.} \label{Tregions}
\centering
\begin{tabular}{l l}
\hline\hline
region & $T$(keV)/$\chi^{2}_{red}$ \\
\hline
&\\
 1 & 6.60$^{+3.14}_{-1.50}$/0.77\\  
 2 & 8.87$^{+0.87}_{-0.69}$/0.95\\ 
 3 & 6.16$^{+1.14}_{-0.87}$/0.69 \\ 
 4 & 10.28$^{+1.64}_{-1.16}$/0.80\\ 
 5 & 9.63$^{+2.47}_{-1.90}$/1.04\\  
 6 & 11.12$^{+1.11}_{-0.74}$/0.90\\ 
 7 & 13.77$^{+3.99}_{-4.11}$/0.66\\
 8 & 9.25$^{+1.09}_{-1.31}$/1.01\\ 
 9 & 7.81$^{+1.01}_{-0.93}$/0.77\\ 
 10 & 12.01$^{+1.45}_{-1.34}$/0.84\\ 
&\\
\hline
\end{tabular}
\end{table}

\subsection{Temperature profiles in different directions}

\begin{figure}[!!htb]
\begin{center}
\resizebox{\hsize}{!}
{
\hspace{0mm}
\includegraphics{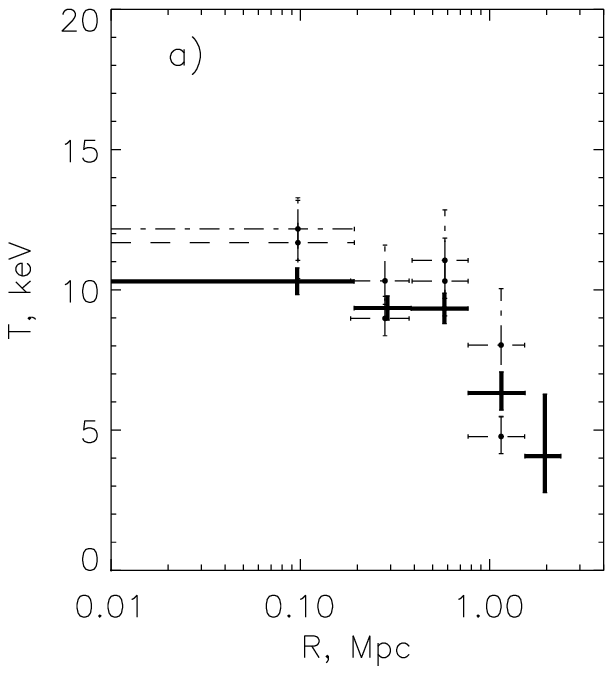}
\hspace{0mm}
\includegraphics{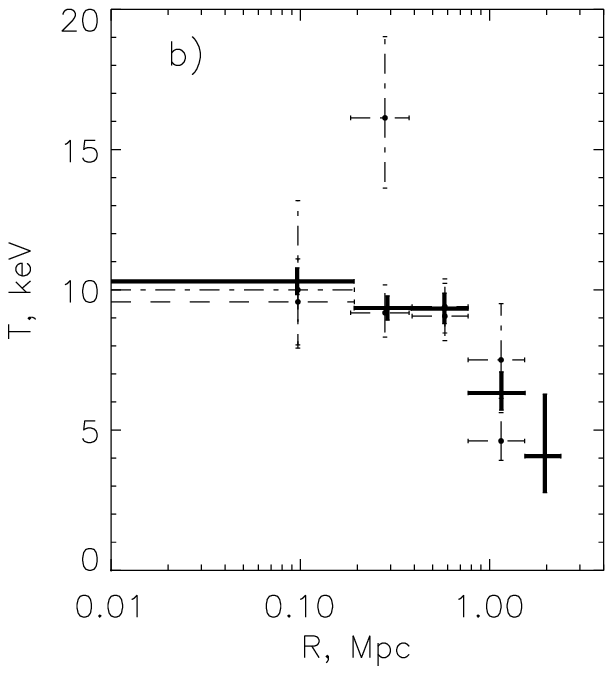}
}
\caption{\label{Tprofiles}The solid points are the total temperature profile. The dotted points are the obtained temperature profiles from sectors: a) the 'hot' temperature profiles ( the SE and NW directions) b) the 'cold' temperature profiles (the SW and NE directions) correspondingly. The 'cold' temperature profile (SW) in panel b) displays a local peak because of the offset of the temperature peak visible in Fig.\ref{TmapReg_Panda.ps}}
\end{center}
\end{figure}

To determine the influence of these variations on the total mass we extracted the temperature profiles in the four different directions. The SE and NW profiles show high temperatures whereas in the SW and NE direction they are lower.  Fig. \ref{TmapReg_Panda.ps} shows the sectors and directions. The reference point of sectors was chosen at the cluster emission centre. The spectra were extracted in each sector up to $4\arcmin$ (1.54 Mpc). The widths of the annuli were chosen similar to that of the total temperature profile.
We obtained the temperature profiles in each direction. Fig. \ref{Tprofiles}a shows the total temperature profile and two 'hot' temperature profiles. Fig. \ref{Tprofiles}b shows the total and 'cold' temperature profiles.

We observed the increase in temperature profile compared with the total temperature profile in the directions of SE and NW and the decrease in SW and NE directions. The maximum temperature variations were observed in the cluster centre. We used these temperature profiles to determine the variation on the total mass profiles.

\section{Mass analysis}

To calculate more precisely the total mass profile up to $R_{200}$ we used the temperature and density profiles obtained above assuming hydrostatic equilibrium. To estimate the influence of the temperature variations on the total mass, we computed the total mass obtained with the different temperature profiles presented in Fig.\ref{massprofile_TmeanTprofile} and corresponding $\beta$- parameters. So we performed a surface brightness analysis in each chosen direction and we obtained $\beta$-model for each profiles (see Table \ref{fit_beta_direction}).

\begin{figure}[!!htb]
\begin{center}\includegraphics[
   width=7.6cm,
   keepaspectratio]{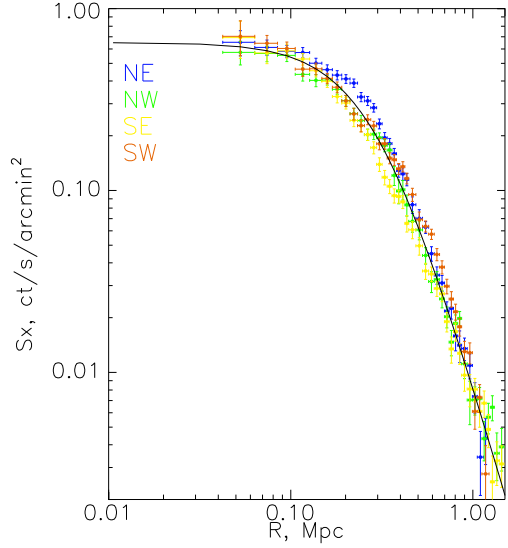}
\caption{\label{art_prof_directions}The surface brightness
profiles in different directions. Solid line is the best fit of the total radial profile.}
\end{center}
\end{figure}

Fig. \ref{art_prof_directions} shows the
obtained surface brightness profiles. We observed
the different perturbation for each direction more significantly in NW and SW
direction from cluster centre to 500 kpc.

\begin{table}[!!htb]
\caption{The $\beta$-model fitting results in different
directions}\label{fit_beta_direction} \centering
\begin{tabular}{l l l l  }
\hline
 \hline
 direction & $\beta$-fit & $r_{c}$(Mpc) & $\chi^{2}_{reduced}$\\
\hline
&\\
   NE &  0.94 $\pm$ 0.01& 0.35 $\pm$ 0.01& 1.05  \\
   SE&  0.65 $\pm$ 0.01  & 0.18 $\pm$ 0.01 & 1.22 \\
   NW &0.72 $\pm$ 0.01 & 0.25 $\pm$ 0.01& 1.68\\
   SW&0.77 $\pm$ 0.01 & 0.31 $\pm$ 0.02& 2.3\\
\hline
\end{tabular}
\end{table}

\subsection{Cluster gas mass}

Using the electron density, we calculated the total gas density of the ICM and the ICM mass by integrating Eq.(3). We obtained a central electron density of $n_{e0} = (7.85\pm0.01\cdot)10^{-3}$cm$^{-3}$ assuming a temperature of 8.85 keV and using  the $\beta$-model parameters obtained from the Nevalainen's background subtraction, given in Table \ref{fit_beta}. The total gas mass is shown in Table \ref{mass}.

\begin{table} [!!htb]
\caption{The results of CL0016+16 mass analysis using different temperature profiles and cluster parameters.} \label{mass}
\begin{tabular}{l l l}
\hline\hline
parameter &M$(<R_{200})$& M$(<R_{500})$ \\
      &$(\times10^{14}M_{\odot})$ & $(\times10^{14}M_{\odot})$ \\
\hline
    $M_{gas}$& $2.24 \pm 0.06$ & 1.41$ \pm $0.04\\
\hline
&\\
    $M_{tot}$,T=8.81 keV& 14.0$ \pm $0.6& 8.4$ \pm $0.3\\
    $M_{tot}$,T(r)& 11.9$ \pm 2.0$&8.0$ \pm 1.0$\\
\hline
&\\
    $M_{tot}$,$T(r)  NW$ hot& 11.2$ \pm $2.6&8.0$ \pm $2.1\\
    $M_{tot}$,$T(r)  NE$ cold& 11.1$ \pm $2.4&8.1$ \pm $1.9\\
    $M_{tot}$,$T(r)  SE$ hot& 14.0$ \pm $2.8&8.5$ \pm $1.8\\
    $M_{tot}$,$T(r)  SW$ cold& 12.0$ \pm $2.5&9.0$ \pm $2.0\\
\hline
&\\
$VT$&&\\
    $M_{tot}$,T = 8.85 keV& 14.42& \\
\hline\hline
&\\
    $n_{e0}(\times10^{-3}$cm$^{-3}$)& 7.85$ \pm $0.01&$\!$\\
    $f_{gas}$& 0.16$ \pm $0.01&\\
    $L_{X}$($\times10^{45}$erg/s) & 5.1$ \pm $ 0.1 & \\
\hline
\end{tabular}
\end{table}

\subsection{Cluster total mass}

Assuming the spherical symmetry and the hydrostatic equilibrium we calculated the gravitational mass of the cluster CL0016+16.
\begin{figure}[!!htb]
\begin{center}\includegraphics[
   width=7.6cm,
   keepaspectratio]{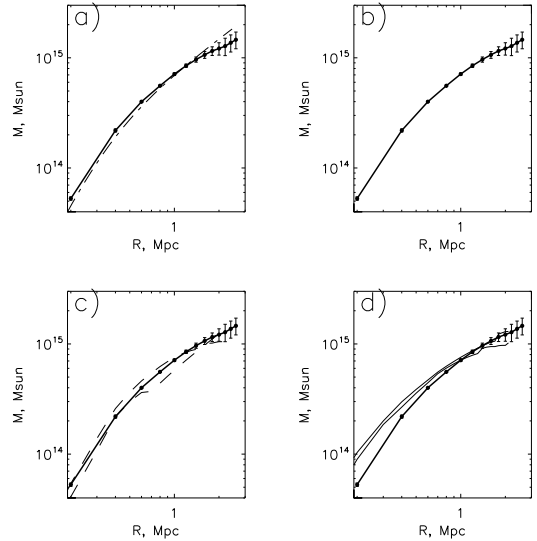}
\caption{\label{massprofile_TmeanTprofile} The integrated total mass profile of CL0016+16. a) The bold solid line, showing the total mass profile, was obtained using the total temperature profile. The dot-dashed line is the total mass profile obtained using constant temperature. b) The bold solid line is the total mass profile obtained using the total temperature profile. c) The solid line is the total mass profile obtained using the 'cold' temperature profiles. d) The dotted line is the total mass profile obtained using the 'hot' temperature profiles.}
\end{center}
\end{figure}

In the first step of the mass analysis we assumed that CL0016+16 is isothermal with the temperature of 8.85 keV and took into account error bars on the temperature profile, $\beta$ and $R_{c}$ parameters. The results are presented in Fig.\ref{massprofile_TmeanTprofile}.

In the second step of the total mass determination we used obtained temperature profiles. Using the hydrostatic equilibrium approach, the $\beta$ parameters obtained with the subtraction Nevalainen's background and the temperature profile we calculated the cluster total mass with:

\begin{equation}
M_{tot}(<r)=-\frac{k}{G\mu m_{p}}r^{2}\left(\frac{dT}{dr}-3\beta T \frac{r}{r^{2}+r_{c}^{2}}\right)
\end{equation}

The mass profile was calculated using the Monte Carlo method, which takes the obtained parameter for the gas density profile and the measured temperature profile as an input. The method of \citet{Neumann} was used, which allows a transformation of the error bars of the temperature profile into error bars of the mass profile.

Table \ref{mass} shows the results of our total mass determination.   Fig.\ref{massprofile_TmeanTprofile} shows the total mass profiles obtained using the constant temperature and assumption temperature profiles.

Using a similar approach of \citet{Kotov} we find the same result on
the total mass. We extrapolated the temperature profile
of \citet{Kotov} to $R_{200}$, obtaining total mass of $(11.7 \pm
1.7)\cdot10^{14}M_{\odot}$, which compare very well with the value of $(11.9 \pm
2.0)\cdot10^{14}M_{\odot}$ using our total temperature profile.

In order to check the influence of temperature variations in the cluster centre on the total mass profiles, we determined four temperature profiles in each direction and the total temperature profile. Using the equilibrium approach and the obtained $\beta$ parameters the total mass was calculated for each temperature profile. The values of the total mass for each temperature profile are shown in the Table \ref{mass} up to $R_{200}$ and $R_{500}$.
 These temperature variations does not significantly affect the total mass, $\Delta M < 20\%$ at the $R_{200}$ in comparison with the total temperature profile. The main contribution to the mass errors in these profiles came from the determination of temperature, which is less constrained in sectors than in the full annuli. It is noticeable anyway that the total mass estimates in sectors are all within 1 $\sigma$ errors.

The total mass profiles for each temperature profile and total temperature profile are shown in Fig.\ref{massprofile_TmeanTprofile}. We found that in the cluster's central parts the total mass profiles obtained with the 'cold' temperature profiles are lower than the total mass profile obtained with the total temperature. At $R_{200}$ we found the predominance of the total mass in the southern part of cluster.

We calculated also the gas mass fraction of 0.16 $\pm$ 0.01 which it is simply the ratio of the ICM mass to the total mass.

\subsection{Total mass from self similarity}

\begin{figure*}[!!htb]
\begin{center}
\resizebox{\hsize}{!}
{
\hspace{0mm}
\includegraphics{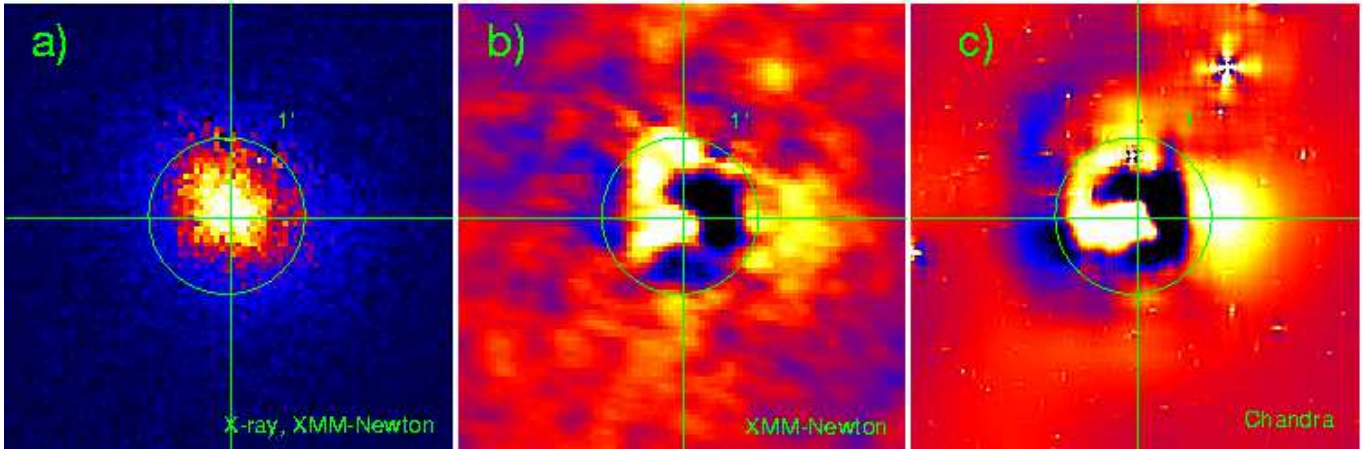}
}
\caption{\label{residuals}a) The X-ray emission from the XMM-Newton data b) The X-ray residual from the XMM-Newton data in the cluster centre, was obtained with the help of 2D $\beta$-model  and c) The obtained X-ray residual from the Chandra data.}
\end{center}
\end{figure*}

Also, the total mass was estimated from $M-T$ scaling relation
using the best fit of the mean temperature in $5\arcmin$ radius at
the cluster redshift, and in $\Lambda$CDM cosmology. Assuming
structural similarity and the virial theorem, we provide a scaling
relation between virial mass, radius and the overall X-ray
temperature $Tx$: $M_{VT200}/R_{VT200}\propto T_{X}$. This
relation corresponds to a fixed density contrast at redshift $z$
and is derived using the virial theorem:
$M_{VT200}/(4/3\pi\rho_{c}R_{VT200}^{3})=200$. This leads to the
well known scaling relation:
\begin{equation}
M_{VT200}\propto(1+z)^{-3/2}T^{3/2}
\end{equation}

To compute the scaling total mass $M_{VT200}$ we used the
normalization factor obtained from numerical simulations of
\citet{Bryan}, rescaled at $R_{VT200}$. The result for $M_{VT200}$
is in the Table \ref{mass}. From the virial theorem Eq.(6)  and
from the hydrostatic equilibrium Eq.(4) we obtained similar
results on the total mass.

\section{Scaling properties}

\subsection{Cluster profile}

\begin{figure}[!!htb]
\begin{center}
\resizebox{\hsize}{!} { \hspace{0mm}
\hspace{0mm}
\includegraphics{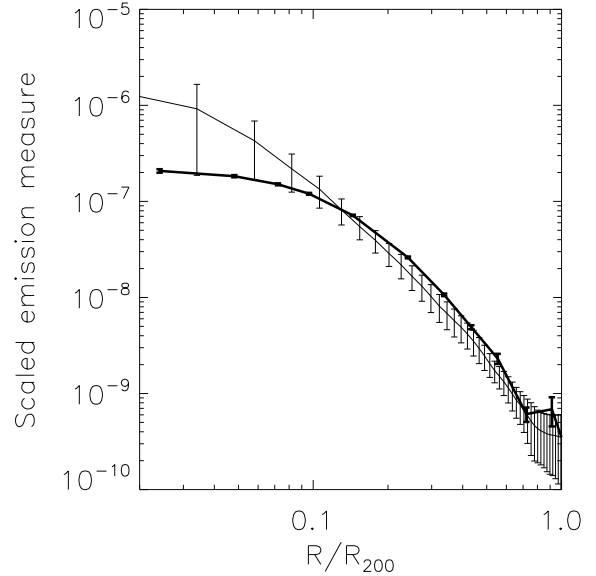}
} \caption{\label{SBscalingR200Reb}a) The dotted lines are the scaled
emission measure (ScEM) profiles of the nearby clusters. The radius
is normalized to $R_{200}$. The bold solid line shows the ScEM
profile of CL0016+16. b) The solid line is the mean dispersion from ScEM
profiles of the nearby clusters. The bold solid line is the SEM
profile of CL0016+16.}
\end{center}
\end{figure}

To study the influence of the cluster properties on the
self-similarity we compared the scaling and structural properties
of distant 'hot' galaxies cluster  CL0016+16 with the scaling
properties obtained from nearby galaxies clusters.

The self similarity model is based on the simple assumptions for
the cluster formation, derived from the top-hat spherical collapse
model \citep{Bryan}. The virialised part of a cluster present at a
given redshift corresponds to a fixed density contrast as compared
to the critical density of the universe at that redshift. This
model makes the definitive predictions in terms of the evolution
of cluster properties.  We considered the scaled emission measure
profiles and the $L_{X}-T$ relation.

We used $R_{200}$ for the normalization on the radius.The
self-similarity of cluster implies that the scaled emission
measure profile (ScEM) should be identical for all clusters:
\begin{equation}
ScEM(r/R_{200})=\frac{4\pi(1+z)^{4}S(r/R_{200})}{\Lambda(T,z)\sqrt{T}(\sqrt{\Delta_{c}}E(z))^{3}}
\end{equation}

$ScEM(r/R_{200})$ were calculated directly from the observed
surface brightness distributions, $S(r)$ is the obtained surface
brightness profile. We applied Eq.(7) to all cluster profiles
using the best fit temperature, corresponding redshift, and the
emissivity $\Lambda(T,z)$.

Using $\Lambda$CDM cosmology, we compared the ScEM profile of CL0016+16
observed with XMM-Newton in the 0.3-2 keV band with
the ScEM profiles of twelve nearby galaxy clusters observed with
ROSAT in the 0.5-2 keV band  \citep{Neumann1}. The nearby galaxy
clusters were obtained in the temperature range ($1.7 < kT < 8.5$)
keV  (but most of them in the 5-6 keV range). The relative error in the calibration of
XMM/EPIC and ROSAT/PSPC can be neglected \citep{Arnaud}.
Fig.\ref{SBscalingR200Reb} shows the average ScEM profiles of
nearby galaxy clusters and the ScEM profile from CL0016+16.

We confirm that, in the cluster centre, the ScEM profiles show a
large dispersion, which are commenly be explained by the non-gravitational
processes. The ScEM profile of CL0016+16 is lower in the centre in
comparison with average ScEM profile, which is the
consequence of the absence of a cooling core. In outer regions
$(r>0.2R_{200})$ the ScEM profile of CL0016+16 is higher  than the
average ScEM profile of nearby galaxy clusters. We compared the total emission from CL0016+16
and from nearby galaxy cluster at the radii
$0.2R_{200}>r>R_{200}$. The total scaled emission of CL0016+16 is
by a factor of 1.6 stronger than the total emission from nearby
galaxy clusters.

Finally, these results can be
explained by cluster properties of CL0016+16, in particular by the
absence of cooling core and high luminosity which in merge activity.


\subsection{$L_{X}-T$ relation}

We tested the scaling $L_{X}-T$ relation for distant luminous
galaxy cluster with possibility of merger. We computed the
bolometric luminosity $L_{X}$ within the $R_{200}$ (see Table
\ref{mass}). The observed count rate in the 0.3-4.5 keV band was
obtained. The integrated surface brightness profile was converted
to bolometric luminosity using the best fit $\beta$-model from
Nevalainen's background subtraction and the instrumental response.
The error in $L_{X}$ includes the statistical errors on the count
rate and temperature. We tested the $L_{X}-T$ using scaling
relation from \citet{Arnaud3}, $L_{X}\propto T^{2.88}$. This
relation was obtained for sample clusters with weak or absent
cooling flow signatures, so it can be used for CL0016+16. Taking
the simple mean temperature, scaling relation by \citet{Arnaud3}
for our cosmology, and normalization evolution factor to our
distant cluster. We found that CL0016+16 is more luminous by a
factor of 2.1 for this temperature. These results confirmed the
possibility of merger in maximum core collapse for CL0016+16,
because the numerical simulation \citep{Randall} suggest that the
effect of the merger is to boost the luminosity and the temperature for
a short time and this can have an obvious effect on scaling relation.

\subsection{2D $\beta$-analysis, X-ray residuals}

To better understand the status of the cluster dynamics in the
centre we performed a two-dimensional fit to the cluster surface
brightness, using a modified $\beta$-model that allows for two
different core radii along the two principal axes of cluster image
ellipse. We fited the surface brightness distribution of CL0016+16
obtained from XMM-Newton data with a 2D $\beta$-model
and we quantified the deviation from this model.

For the XMM-Newton data we used three cameras, with the modelling
and subtraction of background from image (same method were used in
section 3.1.3) and with point sources and gap corrections.

To confirm XMM/Epic 2D  $\beta$ analysis, we used also the
Chandra data to compare the residuals at better spatial resolution.
We are confident of the reality of these residuals since we found
very similar results with XMM/EPIC and Chandra/ACIS.
Reduced cleaned ACIS event list was downloaded directly from Chandra X-ray Center (CXC).
The analysis was performed with CIAO software. The
images were smoothed with Gaussian filter before fitting. We
fitted 2D $\beta$-model in the region $2\arcmin$ (0.8 Mpc
$\approx$ $R_{500}$). The best-fit parameters are listed in Table
\ref{2dbeta}.  All parameters were assumed free in the fit.

The residuals are shown in the Fig. \ref{residuals}. From the
XMM-Newton and Chandra data we obtained very similar results. But
we obtained a different core radius because we used a different
satellite, with a different PSF. Notice, that we do not take the
PSF into account, this explain the different in the core radii.
Two very similar residual maps established the perturbations at
the centre over at least $1\arcmin$ radius ( 0.4 Mpc). In both case
the perturbations in the cluster centre were observed with $20\%$
deviation from the maximum cluster emission. 

After radial projection, we obtained a rather good $\beta$ fit for total surface brightness profile,  but we observe perturbations in 2D. At first glance, CL0016+16 looks rather ÔhomogeneousÕ in brightness and not in temperature. We now, after the 2D  $\beta$ subtraction, enhance the brightness structures in the inner core.
We do observed perturbations in the temperature and in the surface
brightness distributions in the cluster centre. Quantitatively, these 2D  $\beta$- model residuals do not lead to strong variations in density profiles (thus in the mass estimate in the HE scheme), but they are of importance to qualify the relaxation status of the inner core of CL0016+16 and thus the limit of the hydrostatic equilibrium hypothesis.

If CL0016 is at the maximum core collapse phase, it will present very similar residuals as observed in numerical simulations around this phase. See, for example, figure 4 in \citet{Sarazin}, where they present a 1 to 1 mass ratio collision with high impact parameter. Following that scheme, we can estimate the time since core collapse passage by more precise comparison with these simulations.  CL0016+16 is 2 time too luminous as what the simulation found for a cluster at 0.01 Gyr around its maximum core collapse (see figure 5 in \citep{Sarazin}).


\begin{table} [!!htb]
\caption{The best-fitting results of the 2D $\beta$-model, from the XMM-Newton and Chandra data}
\label{2dbeta} \centering
\begin{tabular}{l l l}
\hline\hline
Parameter& Best-fitting \\
& XMM & Chandra\\
\hline
&\\
$R_{c1}$(Mpc)& 0.291 & 0.237 \\
$R_{c2}$(Mpc) & 0.364 & 0.295\\
$\beta$ & 0.79 & 0.79\\
Pa & 2.24 & 2.33\\
RA & 00:18:33 & 00:18:33\\
Dec & 16:26:06& 16:26:11\\
&\\
\hline
\end{tabular}
\end{table}

\section{Discussion}

In previous works, CL0016+16 was always considered as a very massive {\it relaxed} cluster.
In our present analysis, we found the strong evidence that, at least in the centre, the cluster is not relaxed.
From the weakest to the strongest argument favouring the merger scenario, we found: \\
\begin{itemize}
\item CL0016+16 was about a factor of 2 too bright, when compared to the expected luminosity of the $L_{X}-T$ relation.\\
\item We detected similar significant residuals after subtraction of a 2D $\beta$-model fit using both XMM-Newton and Chandra observations, especially in the inner region ($r<1\arcmin \approx 400$ kpc).
Notice that the X-ray maximum was not onto the centre of the D $\beta$-model fit.\\
\item The temperature map of the $2.5\arcmin$ region (i.e. up to 1 Mpc) clearly looks like an equal mass merger at first maximum core collapse showing significant azimuthal variations.\\
\end{itemize}

In addition, CL0016+16 is also well known to host a strong radio halo which also argues in favour of a merging cluster \citep{Feretti}.
Also we obtained a high velocity distribution of galaxies for this cluster.
Using NED data, we obtained a  $\sigma_{v} = 1800$ km s$^{-1}$ for the 150 galaxies at CL0016+16 redshift. The maximum of this map is offset by $0.4\arcmin$ (141 kpc) with the X-ray one.
We tried to quantify the impact this could have onto the total mass estimation and found that depending on the adopted temperature profile.
(but always in the framework of Hydrostatic
Equilibrium), the total mass may vary by something like $20$\%, which is much bigger than the quoted error which are around 5\%
(the errors in the temperature profile is the main contributor to this error budget).

%
The main results of our detailed analysis of the galaxy cluster CL0016+16 from the
XMM-Newton data can be summarized as follows.

We performed a detailed image and spectral analysis of the
XMM-Newton data using three different backgrounds. In all cases we
detected the cluster emission up to $R_{200}$ with the XMM-Newton
data. Also using the template of simulated cluster we checked the
possibility of detection with XMM-Newton data up to  $R_{200}$ (see
Appendix \ref{appendix}).
All obtained surface brightness profiles give a good fit with the $\beta$-model.

We studied the dynamics of CL0016+16 from the detailed spectral
analysis. The global temperature estimated in 5$\arcmin$ is 8.81
keV, but the spectral study shows that this cluster is not
isothermal. We obtained the temperature profile which decreases in
the outer regions, our results are in a good agreement with results
from \citet{Kotov}. The temperature map shows the asymmetry in the
radius $1\arcmin$, the temperature maximum (T = 12.0 $\pm$ 1.5 keV)
is not superposed on the centre X-ray cluster emission and is located
to south-western direction; also we observed the cold regions (T =
6.2$\pm$ 1.2 keV) to north-east. To study the dynamics of
CL0016+16 more precisely, we chose several regions using the temperature map. We
extracted temperature in the regions, found no
spherical symmetry in temperature and the presence of the 'cold' and
'hot' regions in the cluster centre.

We tested the influence of the temperature variations on the cluster
total mass using the obtained 'cold' and 'hot' temperature profiles. We
calculated total mass within $R_{200}$ and $R_{500}$, the
temperature variations do not significantly impact on the total
mass, $\Delta M<20\%$ at $R_{200}$. But the temperature variations
in the centre influence mainly on the inner part of total mass profile. To
better understand the possible dynamics in the cluster centre we
calculated 2D $\beta$-model from XMM-Newton and Chandra observations. The
same residuals were obtained in the cluster centre, which may be an
indication of merging.

We tested CL0016+16 on the self similarity theory.  ScEM of distant galaxy cluster CL0016+16 is higher than the mean ScEM profile of nearby cluster in external regions. In the centre we saw larger dispersion, which can be explained by non-gravitational processes. We tested  $L_{X}-T$ relation using scaling relation from \citet{Arnaud3} and concluded that CL0016+16 is more luminous by factor  of 2 for this temperature.
It is also the argument for maximum core collapse in CL0016+16.

\begin{acknowledgements}
We would like to thank Dominique Aubert for performing the numerical simulations of Cluster 6 and making the simulated data available
for us to test the possibility of detection up to $R_{200}$ with XMM-Newton. We also would like to thank Herve Bourdin for making
available to us the XWSM code.
\end{acknowledgements}

\appendix
\section{Possibility of detection up to $R_{200}$ with XMM-Newton}
\label{appendix}

In order to check the possibility of the cluster emission detection
up to $R_{200}$ with XMM-Newton data we decided to treat a template of similar
cluster from the Hydro N-body cosmology simulation, code RAMSES
\citep{Teyssier}.

The leading idea to use numerical simulation was only to check the ability of XMM/EPIC to detect such a massive cluster at this redshift up to its virial radius taking as precisely as possible all the observational effects into account in particular, the relationship between the background, the instrumental response and the hot temperature of this cluster. The scaling applied onto Cluster6 directly follow the M-T relation and allow to properly study the instrumental effect and limits.

We want to obtain similar cluster using the adaptive mesh refinement 3-D hydrodynamical cosmology simulation of structure formation \citep{Teyssier}. Cluster 6 was formed at a temperature of 3.4 keV.

Using the self-similarity theory from Cluster 6 we scaled its temperature and mass to those of CL0016+16 without changing the radial profile (below, we refer to this cluster as simulated cluster). To obtain the cluster emission from the simulation, we created the photons using the Monte Carlo approach. We calculated emission measure per cell using the cell size, density and the solid angle which depends on the used cosmology \citep{Bourdin}.

\begin{figure}[!!htb]
\begin{center}\includegraphics[
   width=7.6 cm,
   keepaspectratio]{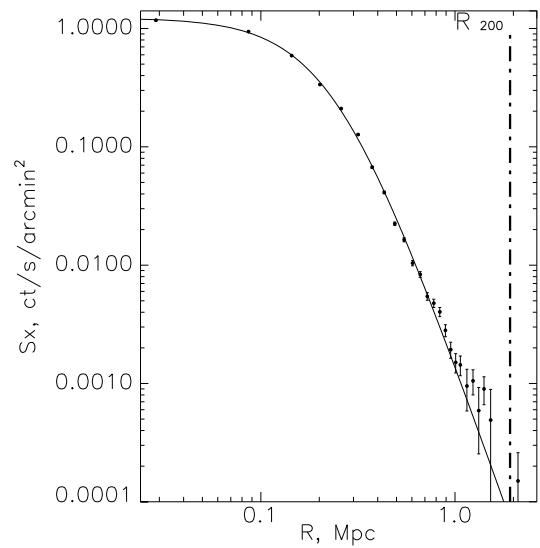}\end{center}
\caption{\label{art_prof_SimuAmas6_RPsoustrait} The surface brightness profile for similar cluster of CL0016+16 obtained from the Hydro N-body cosmology simulations.}
\end{figure}

The photons were obtained from cosmology simulation for the simulated cluster. We convolved these photons with XMM-Newton response. In our analysis we took into account the PSF effect, vignetting correction function,  mask of CCD cameras (gap) and added the randomly selected events from background data of A. Read. The same data reduction was performed as for CL0016+16. Fig.4 shows the resulting surface brightness profile for the simulated cluster.  Notice that Cluster 6 is more cuspy than CL0016+16, its $\beta$-model parameters are $\beta$ = 0.99 and the core radius $r_{c}$ = 0.24 Mpc. We still detect the emission of simulated cluster up to $R_{200}$ but with less significance, see Fig.\ref{art_prof_SimuAmas6_RPsoustrait}. This is a rather robust confirmation that XMM-Newton can detect the bright cluster emission up to $R_{200}$.

\begin{thebibliography}{}

    \bibitem[Arnaud et al., 2002]{Arnaud} Arnaud, M., Majerovwicz, S., Lumb D., et al, 2002,
     A\&A, 390, 27

    \bibitem[Arnaud et al., 2001]{Arnaud1} Arnaud, M., Neumann, D.M., Aghanim, N., et al, 2001,
     A\&A, 365, L80

    \bibitem[Arnaud et al., 2002]{Arnaud2} Arnaud, M., Aghanim, N., Neumann, D.M., 2002,
     A\&A, 389, 1

    \bibitem[Arnaud and Evrard, 1999]{Arnaud3} Arnaud, M. and  Evrard, E.A., 1999,
     MNRAS, 305, 631

     \bibitem[Baumgartner et al., 2005]{Baumgartner} Baumgartner, W. H., Loewenstein, M., Horner, D. J., \& Mushotzky, R. F., 2005,
     ApJ, 620, 680

     \bibitem[Birkinshaw et al., 1981]{Birkinshaw} Birkinshaw, M., Gull, S.F. and Moffet, A.T., 1981,
     ApJ, 251, L69

    \bibitem[Belsole et al., 2005]{Belsole} Belsole, E., Sauvageot, J.L., Pratt, G.W., \& Bourdin H., 2005,
     A\&A, 430, 385

    \bibitem[Belsole et al., 2004]{Belsole1} Belsole E., Pratt, G.W., Sauvageot, J.L., \& Bourdin H., 2004,
     A\&A, 415, 821

    \bibitem[Bourdin et al., 2004]{Bourdin} Bourdin, H., Sauvageot, J.L., Slezak, E., et al, 2004,
     A\&A, 414, 429

    \bibitem[Bonamente et al., 2006]{Bonamente} Bonamente, M.,  Joy, M., La Roque, S., et al, 2006,
     ApJ, 647, 25

    \bibitem[Bryan \& Norman, 1998]{Bryan} Bryan, G.L. \& Norman M.L., 1998,
     A\&A, 495, 80

    \bibitem[Cavaliere and Fusco-Femiano, 1976]{Cavaliere} Cavaliere, A., Fusco-Femiano, R., 1976,
     A\&A, 49, 137

    \bibitem[Clowe et al., 2000]{Clowe} Clowe, D., Luppino, G.A., Kaiser, N. and Cioia, M.I., 2000,
     ApJ, 539, 540

    \bibitem[Comerford et al., 2006]{Comerford} Comerford, J.M., Menegnetti, M.,Bartelmann, M., \& Schirmer, M., 2006,
     ApJ, 642,39

    \bibitem[Cole \& Lacey, 1996]{Cole} Cole, S.,\& Lacey, C., 1996,
     MNRAS, 281,716

    \bibitem[Evrard et al., 1996]{Evrard} Evrard, A.E., Metzler, C.A., Navarro, J.F., 1996,
     ApJ, 469, 494

    \bibitem[Giovannini \& Feretti, 2000]{Feretti} Giovannini, G., Feretti, L., 2000,
     New Astronomy 5, 335

    \bibitem[Hughes \& Birkinshaw, 1998]{Hughes} Hughes, J.P. \& Birkinshaw, M., 1998,
     ApJ, 497, 645

    \bibitem[Kotov \& Vikhlinin, 2005]{Kotov} Kotov, O. \& Vikhlinin, A., 2005,
     ApJ, 633, 781

    \bibitem[Majerowicz et al., 2002]{Majerowicz} Majerowicz, S., Neumann D. \& Reiprich, T., 2002,
     A\&A, 394, 77

    \bibitem[Majerowicz et al., 2004]{Majerowicz1} Majerowicz, S., Neumann, D.M., Romer, A.K., et al, 2004,
     A\&A, 444, 673

    \bibitem[Neumann \& Bohringer, 1997]{Neumann}  Neumann, D.M. \& Bohringer H., 1997,
     MNRAS, 289, 123

    \bibitem[Neumann, 2005]{Neumann1} Neumann, D., 2005,
     A\&A, 439, 465

    \bibitem[Neumann \& Arnaud, 1999]{Neumann2} Neumann, D. \& Arnaud M., 1999,
     A\&A, 348, 711

    \bibitem[Nevalainen et al., 2005]{Nevalainen} Nevalainen, J., Markevitch, M., Lumb, D., 2005,
     ApJ, 629, 172

     \bibitem[Randall et al., 2002]{Randall} Randall, S.W., Sarazin, and G.L., Ricker, P.M., 2002,
     ApJ, 577, 579

    \bibitem[Read \& Ponman, 2003]{Read} Read, A. \& Ponman, T., 2003,
     A\&A, 409, 395

    \bibitem[Ricker and Sarazin, 2001]{Sarazin} Ricker, P.M. and Sarazin, G.L., 2001,
     ApJ, 561, 621

    \bibitem[Sauvageot et al., 2005]{Sauvageot} Sauvageot, J.L., Belsole, E., \& Pratt, G.W., 2005,
     A\&A, 444, 673

    \bibitem[Vikhlinin et al., 1999]{Vikhlinin} Vikhlinin, A., Forman W., Jones, C., 1999,
     ApJ, 525, 47

    \bibitem[Tanaka et al., 2005]{Tanaka} Tanaka, M., Kodama,T., Arimoto, N., et al, 2005,
     MNRAS, 362, 268

    \bibitem[Teyssier, 2002]{Teyssier} Teyssier, R., 2002,
     A\&A 385, 337

    \bibitem[Worrall \& Birkinshaw, 2003]{Worrall}  Worrall, D.M. \& Birkinshaw, M., 2003,
     MNRAS, 340, 1261

\end{thebibliography}
\end{document}